\documentclass[aps,prl,
twocolumn, superscriptaddress,showpacs,floatfix,nobibnotes]{revtex4}

\usepackage{epsfig}
\usepackage{epstopdf}
\usepackage{color}
\usepackage{float}
\usepackage{graphicx}
\usepackage{longtable}
\usepackage{CJK}

\begin{document}

\title{Exact solutions of an extended Bose-Habbard model with  $E_2$ symmetry}

\begin{CJK*}{GBK}{song}

\author{Feng Pan}
\affiliation{Department of Physics, Liaoning Normal University,
Dalian 116029, China}\affiliation{Department of Physics and
Astronomy, Louisiana State University, Baton Rouge, LA 70803-4001,
USA}

\author{Ningyun Zhang}
\affiliation{Department of Physics, Liaoning Normal University,
Dalian 116029, China}

\author{Qianyun Wang}
\affiliation{Department of Physics, Liaoning Normal University,
Dalian 116029, China}

\author{J. P. Draayer}
\affiliation{Department of Physics and Astronomy, Louisiana State
University, Baton Rouge, LA 70803-4001, USA}
\date{\today}

\begin{abstract}
An extended Bose-Hubbard (BH) model with number-dependent multi-site and  infinite-range hopping
is proposed, which, similar to the original BH model, describes a phase transition between the delocalized
superfluid (SF) phase and localized Mott insulator (MI) phase.
It is shown that this extended model with local Euclidean $E_{2}$ symmetry is exactly solvable
when on-site local potential or disorder is included,
while the model without local potential or disorder is quasi-exactly solvable, which means only a part of the
excited states including the ground state being exactly solvable.
As applications of the exact solution for the ground state,
phase diagram of the model in 1D without local potential and on-site disorder for filling factor $\rho=1$
with  $M=6$ sites and that with $M=10$  are obtained.
The probabilities to detect $n$ particles on a single site, $P_{n}$, for $n=0,~1,~2$  as functions
of the control parameter $U/t$  in these two cases are also calculated. It is shown that
the critical point in $P_{n}$
and in the entanglement measure is away from that of the SF-MI transition determined in the phase analysis.
It is also shown that the the model-independent entanglement measure
is related with $P_{n}$, which, therefore, may be practically useful
because $P_{n}$ is measurable experimentally.

\end{abstract}


\maketitle

\end{CJK*}

{\bf Introduction.} As is well known, the Bose-Hubbard (BH) model originally introduced by Fisher et al \cite{[1]}
provides a reasonable description of experiments of ultracold bosons in
an optical lattice, in which the boson hopping strength and on-site
interaction can easily  be tuned by controlling the laser intensity
and by means of Feshbach resonances~\cite{[2],[3],[4]}.
Generally speaking, the model is non-integrable for more than two sites.
Owing to the fact that its numerical diagonalization can only be carried out for small systems due to
the enormous size of the Hilbert subspace with dimension $D(N,M)=(N+M-1)!/(N!(M-1)!)$,
where $N$ and $M$ stand for the number of bosons and the number of lattice sites,
respectively, perturbation theory~\cite{[1],[5],[6],[7]}, quantum Monte-Carlo (QMC) calculations~\cite{[8],[9]},
the density matrix renormalization group (DMRG) method~\cite{[10],[11],[12]} were used, from which
ground state properties of the model have been
studied extensively. Extensions of the model to include on-site disorder~\cite{[1],[13],[14]},
longer ranged interactions~\cite{[5],[15]}, long-range hopping~\cite{[16],[17]}, infinite-range hopping~~\cite{[1],[18]},
and pair-correlated hopping~\cite{[19]}, etc.
have also been made, in which, generally, more complicated phase structures emerge.
Though many properties of the BH model have been known quite well from the above mentioned approximate
calculations, it will be helpful if there is a similar model
that can be solved exactly or quasi-exactly,
because exactly and
quasi-exactly solvable models may offer valuable insight
and their solutions may be used as the basis in approximation methods.

{\bf The extended Bose-Hubbard model.}
Similar to the original BH model~\cite{[1]}, we consider an extended Bose-Hubbard Hamiltonian
with number-dependent multi-site and infinite-range hopping terms, of which the
Hamiltonian may be written as

$$\hat{H} =-t_{0}\sum_{k=1}^{\infty}\sum_{j_{1}\leq\cdots\leq j_{k}}\tilde{b}_{j_{1}}^{\dagger}\cdots \tilde{b}_{j_{k}}^{\dagger}
\sum_{j^{\prime}_{1}\leq\cdots\leq j^{\prime}_{k}}\tilde{b}_{j_{1}^{\prime}}\cdots \tilde{b}_{j^{\prime}_{k}}
+$$
\begin{equation}{~U\over{2}}\sum_{j} \hat{n}_{j}(\hat{n}_{j}-1)+\sum_{j}\epsilon_{j}\hat{n}_{j},
\end{equation}
where $\{\epsilon_{j}\}$ are the local potentials including effective random local shifts representing the disorder,
$t_{0}$ and $U$ are real parameters,
and two groups of site-indices $\{j_{1},\cdots,j_{k}\}$
and $\{j^{\prime}_{1},\cdots,j^{\prime}_{k}\}$ in the restricted sum of
the first term of (1) run over all sites
with the restriction that no any one of $\{j_{1},\cdots,j_{k}\}$
equals to any one of $\{j^{\prime}_{1},\cdots,j^{\prime}_{k}\}$,
but the site-indices in the same group can be taken as the same,
which describes
bosons hopping from the sites  $\{j^{\prime}_{1},\cdots,j^{\prime}_{k}\}$ to
the sites $\{j_{1},\cdots,j_{k}\}$ simultaneously. In (1),
$\tilde{b}_{j}^{\dagger}={b}_{j}^{\dagger}f(\hat{n}_{j})$ and $\tilde{b}_{j}=f(\hat{n}_{j})b_{j}$, where
$f(\hat{n}_{j})=1/\sqrt{\hat{n}_{j}+1}$ is  well-defined functional of the
local boson number operator on site $j$ with $\hat{n}_{j}=b^{\dagger}_{j}b_{j}$, and
$b_{j}^{\dagger}$ ($b_{j}$) is the usual boson creation (annihilation) operator.
Obviously, the operators $\{\tilde{b}_{j},~\tilde{b}^{\dagger}_{j},~
\hat{n}_{j}\}$ generate the two-dimensional Euclidean algebra $E_{2}$, which satisfy the
commutation relations:

\begin{equation}
[\hat{n}_{j},~\tilde{b}_{j}]=-\tilde{b}_{j},~
[\hat{n}_{j},~\tilde{b}^{\dagger}_{j}]=\tilde{b}^{\dagger}_{j},~
[\tilde{b}_{j},~\tilde{b}^{\dagger}_{j}]=0.
\end{equation}

Though the extended model can also be studied in high dimensional cases,
only one-dimensional $M$-site case will be considered in the following.
Rewriting the deformed boson operators $\{\tilde{b}_{j},~\tilde{b}^{\dagger}_{j}\}$
in terms of the usual boson operators, one may observe that (1) is equivalent to
a Bose-Habbard model with number-dependent multi-site and infinite-range hopping.
For example, in the two-site (dimer) case, the first term of the Hamiltonian (1) becomes
$\hat{H}^{\rm hop}_{\rm Dimer} =-\sum_{k=1}^{\infty}
\left(t^{(k)}_{12}{b}^{\dagger k}_{1}{b}^{k}_{2}
+t^{k}_{21}{b}^{\dagger k}_{2}{b}^{k}_{1}\right)$,
where
$t^{(k)}_{ij}=t_{0}\sqrt{\hat{n}_{i}!(\hat{n}_{j}-k)!\over{\hat{n}_{j}!(\hat{n}_{i}+k)!}}$,
which shows that the hopping matrix elements depend on the number of bosons
in the initial and the target sites.
The more the number of bosons on the two sites, the less the hopping
strength.

In order to study this exaggerated hopping situation,
we consider the Hamiltonian (1) with $U=0$ and $\epsilon_{j}=0~\forall~j$. It can easily be
shown that the ground state of (1) in this case is non-degenerate and also
most coherently delocalized with

\begin{equation}\label{SF-gs}
\vert N\rangle_{\rm g}={\cal N}\sum_{n_{1},\cdots,n_{M}}\vert n_{1},\cdots,n_{M}\rangle,
\end{equation}
where $\{\vert n_{1},\cdots,n_{M}\rangle=\tilde{b}_{1}^{\dagger n_{1}}\cdots \tilde{b}_{M}^{\dagger n_{M}}\vert 0\rangle\}$
are boson Fock states, in which $\vert 0\rangle$ is the boson vacuum state,
the sums in (\ref{SF-gs}) run over all possible positive integer values with restriction
$\sum_{i=1}^{M}n_{i}=N$ because the total number of bosons in the model
is a conserved quantity, and ${\cal N}^{-2}$ is simply equal to the dimension
of the Hilbert subspace with ${\cal N}^{-2}=D(N,M)$.
The corresponding ground state energy is given by $E_{\rm g}^{\rm SF}=-t_{0}(D(N,M)-1)$.
While other excited states are all degenerate with excitation energy being zero.
In contrast to the superfluid (SF) phase in the original BH model, in which
there are gapless quasi-particle excitations, in our extended model,
however, there is the energy gap $\Delta=t_{0}(D(N,M)-1)$ in the spectrum. As a consequence,
bosons in this extended  model is not easily excitable. Moreover,
in the original BH model, the ground state energy in this case is given by
$E_{\rm g}\sim -2tN$, where $t$ is the nearest neighbor hopping strength.
In comparison to the original BH model,
one finds that $t_{0}\sim 2tN/D(N,M)$
in order to reproduce
the same ground state energy as that in the original BH model.
Hence, we set $t_{0}=2tN/D(N,M)$, where $t$ is used as an alternative parameter
in the extended model.

Next, we show the ground state probability
to detect $n$ particles on a given lattice site, $P_{n}$,
defined as

\begin{equation}
P_{n}=\sum_{n_{i\neq 1}}\vert\langle n_{1}=n,\{n_{i\neq 1}\}\vert N\rangle_{\rm g}\vert^2
\end{equation}
at a fixed filling factor $\rho=N/M$ for $U=0$ and $\epsilon_{j}=0~\forall~j$.
It is well known that $P_{n}$ of the original BH model in this case
obeys the Poisson distribution  with

\begin{equation}\label{Poisson}
P_{n}=e^{-\rho}{~\rho^{n}\over{n!}}.
\end{equation}
In the extended BH model with $U=0$ and $\epsilon_{j}=0~\forall~j$, by
using the explicit expression
of the ground state (\ref{SF-gs}), it can easily be obtained that

\begin{equation}\label{frac}
P_{n}={D(N-n,M-1)\over{D(N,M)}}.
\end{equation}
As shown in Fig. 1, there is little difference in the two distributions
(\ref{Poisson}) and (\ref{frac}) when $\rho$ is small, especially when $\rho<1$ and for large $n$ cases.
With the increasing of the filling factor $\rho$, there is small deviation from the Poisson distribution
in $P_{n}$ in the extended model when $n$ is small. Therefore, though the boson hopping is exaggerated
in the extended model in comparison to the original BH model, to some extent, the Hamiltonian (1)
does describe a phase transition from SF to Mott insulator (MI) governed by the competition
of the boson mobility and on-site interaction similar to that occurs in the original BH model.

\begin{figure}[H]
\centerline{\hbox{\epsfig{file=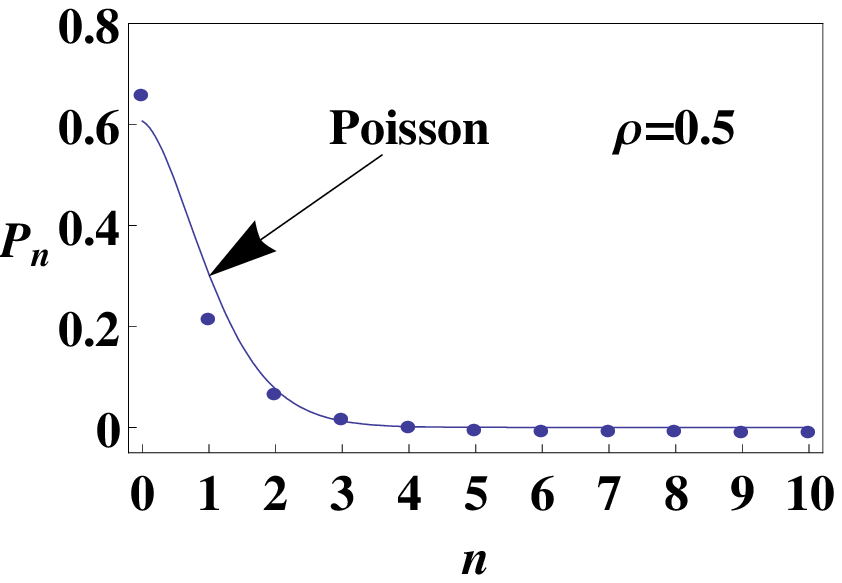, width=4.1cm}~\epsfig{file=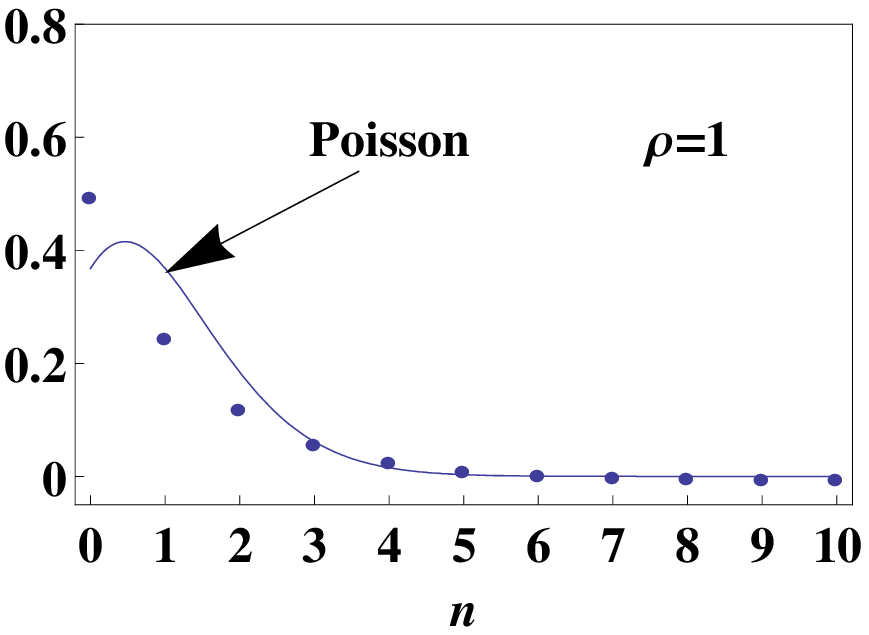, width=3.85cm}}}
\centerline{\hbox{~\epsfig{file=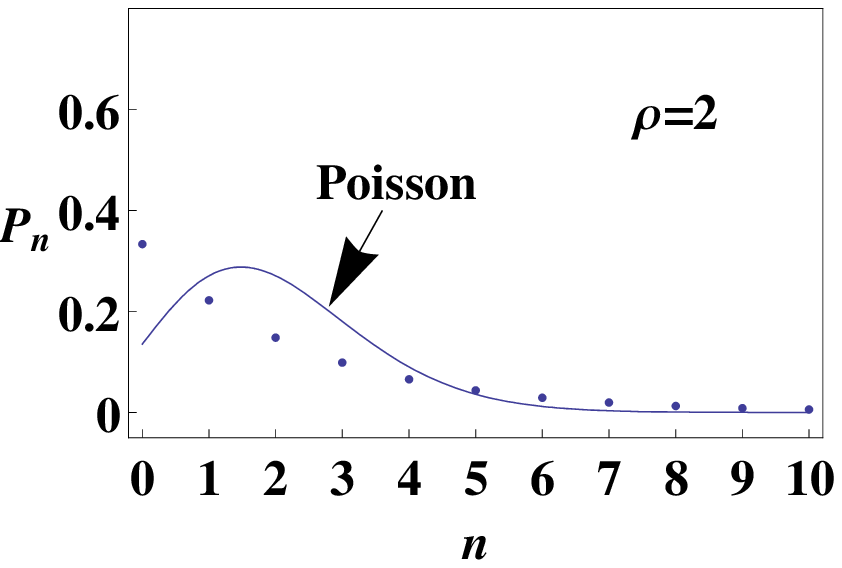, width=4.1cm}~\epsfig{file=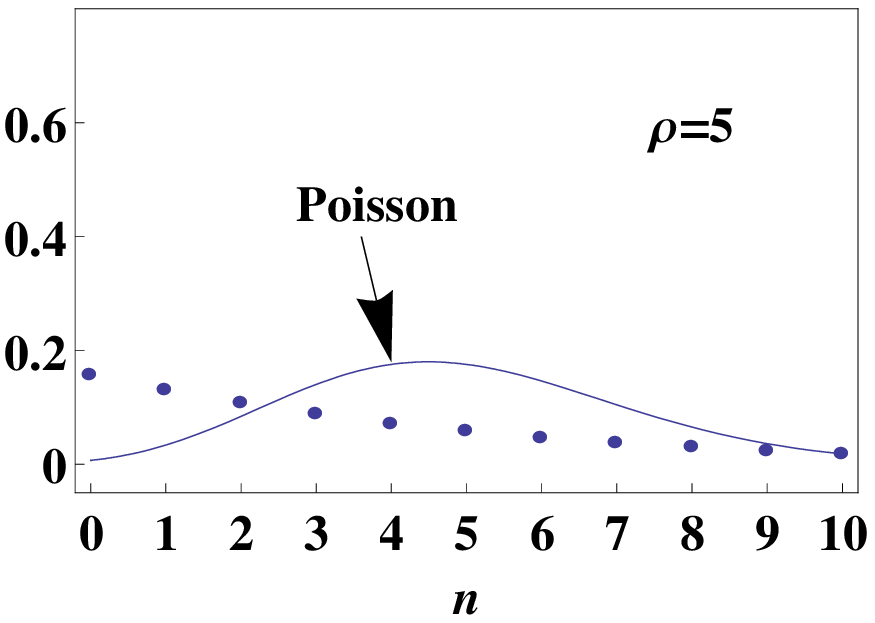, width=3.9cm}}}
\caption{(Color online) The ground state probability
to detect $n$ particles on a given lattice site, $P_{n}$, at $U=0$ with $\epsilon_{j}=0~\forall~j$ and
fixed filling factors $\rho=0.5,~1,~2,$ and $5$, respectively,
 in the extended BH model and compared to that obeying the Poisson distribution (\ref{Poisson}) in the
original BH model. The calculations were carried out with $M=10^5$ sites.}
\label{even-odd}
\end{figure}

The advantage of including the number-dependent multi-site and infinite-range hopping terms in replace
of the nearest neighbor hopping term lies in the fact that the Hamiltonian  (1)
is exactly solvable when $\epsilon_{j}$ ($j=1,\cdots, M)$ are not equal one another, and is
quasi-exactly solvable when some of $\epsilon_{j}$ ($j=1,\cdots, M)$ are the same.
To digonalize the Hamiltonian (1) for $N$ particles on an $M$-site lattice,
we use the simple algebraic Bethe ansatz with

\begin{equation}\label{BA}
\vert N,\zeta\rangle=\sum_{n_{1},\cdots,n_{M}}C^{(\zeta)}_{n_{1},\cdots,n_{M}}
\vert n_{{1}},\cdots,n_{{M}}\rangle,
\end{equation}
where the sum is restricted with $\sum_{j=1}^{M}n_{j}=N$,
and $C^{(\zeta)}_{n_{1},\cdots,n_{M}}$ is the expansion coefficient to be determined.
Similar to the procedures used in~\cite{[20],[21]}, it can be proven that the expansion
coefficient  $C^{(\zeta)}_{n_{1},\cdots,n_{M}}$ for $U\neq 0$ can be expressed as

\begin{equation}\label{BAC}
C^{(\zeta)}_{n_{1},\cdots,n_{M}}={{1}\over{F(n_{1},\cdots, n_{M})}},
\end{equation}
where $F(n_{1},\cdots, n_{M})=E^{(\zeta)}/U-t_{0}/U-{1\over{2}}\sum_{j=1}^{M}n_{j}(n_{j}-1)-
\sum_{j=1}^{M}(\epsilon_{j}/U)n_{j}$, in which
$E^{(\zeta)}$ is the $\zeta$-th eigen-energy. To show that  (\ref{BA}) and (\ref{BAC}) are indeed
consistent, one may directly apply the Hamiltonian (1) on the $N$-particle state (\ref{BA}) to establish
the eigen-equation $\hat{H}\vert N,\zeta\rangle=E^{(\zeta)}\vert N,\zeta\rangle$.
Since the two groups of site-indices $\{j_{1},\cdots,j_{k}\}$
and $\{j^{\prime}_{1},\cdots,j^{\prime}_{k}\}$ in the restricted
sum of the hopping term of (1) run over all sites
with the restriction that no any one of $\{j_{1},\cdots,j_{k}\}$
equals to any one of $\{j^{\prime}_{1},\cdots,j^{\prime}_{k}\}$,
after simple algebraic manipulation, one can easily find that

$${\small -t_{0}\sum_{k=1}^{\infty}\sum_{j_{1}\leq\cdots\leq j_{k}}\tilde{b}_{j_{1}}^{\dagger}\cdots \tilde{b}_{j_{k}}^{\dagger}
\sum_{j^{\prime}_{1}\leq\cdots\leq j^{\prime}_{k}}\tilde{b}_{j_{1}^{\prime}}\cdots \tilde{b}_{j^{\prime}_{k}}\vert N,\zeta\rangle=}$$
\begin{equation}\small
\small t_{0}\vert N,\zeta\rangle-t_{0}\sum_{n^{\prime}_{1},\cdots,n^{\prime}_{M}}C^{(\zeta)}_{n^{\prime}_{1},\cdots,n^{\prime}_{M}}\sum_{n_{1},\cdots,n_{M}}
\vert n_{{1}},\cdots,n_{{M}}\rangle.
\end{equation}
Once the expansion coefficient is chosen as that shown in (\ref{BAC}),
the eigen-equation $\hat{H}\vert N,\zeta\rangle=E^{(\zeta)}\vert N,\zeta\rangle$
is fulfilled when and only when


\begin{equation}\label{BAE}
-(t_{0}/U)\sum_{n_{1},\cdots,n_{M}}{1\over{F(n_{1},\cdots, n_{M})}}
=1.
\end{equation}
Solutions of (\ref{BAE}) provide with eigenvalues $E^{(\zeta)}$ and
 the corresponding eigenstates (\ref{BA})
simultaneously.

When $\epsilon_{j}$ ($j=1,\cdots, M)$ are not equal one another,
binomials  $F(n_{1},\cdots, n_{M})$ with variable $E^{(\zeta)}$ in the denominators
of terms in the sum of (\ref{BAE}) are all different. Therefore, (\ref{BAE}) in this case results in
a polynomial equation with variable $E^{(\zeta)}$. The degree of the polynomial  equals exactly to
the dimension of the concerned Hilbert subspace $D(N,M)$. There are exactly
$D(N,M)$ distinct roots $E^{(\zeta)}$ of (\ref{BAE}) in this case. Hence, the extended
BH Hamiltonian (1) in this case is exactly solved.
When some of $\epsilon_{j}$ ($j=1,\cdots, M)$ are the same,
the number of distinct terms in the sum of (\ref{BAE}) will decrease,
especially when local potential and on-site disorder are neglected
with $\epsilon_{j}=0~\forall~j$. In this case,
generally, (\ref{BAE})
only provides a part of solutions for (1), which is thus called
quasi-exactly solvable. Most importantly, roots of (\ref{BAE})
always include the lowest eigenvalue of (1) in the
full matrix diagonalization
even when
$\epsilon_{j}=0~\forall~j$, which is mainly due to
the site-permutation group $S_M$ symmetry of (1)
in this case.
Firstly, the on-site repulsion term in
(1) keeps the same contribution to the energy for symmetric,
anti-symmetric, and mixed representations of $S_M$,
which can easily be verified when it is directly diagonalized
within the concerned Hilbert subspace.
As is shown in (\ref{SF-gs}), the ground state
in the SF phase with $U=0$ is always symmetric.
Only excited states in the SF phase may be non-symmetric.
Therefore, the ground state of (1) is always
symmetric with respect to the site-permutation.
It is obvious that the Bethe ansatz eigenstates shown by (\ref{BA}) and (\ref{BAC})
are always symmetric with respect to the site-permutation
when $\epsilon_{j}=0~\forall~j$, which ensures roots obtained from (\ref{BAE})
involving that corresponding to the ground state of the system in this case.

\begin{figure}[H]
\begin{center}
\includegraphics[scale=0.46]{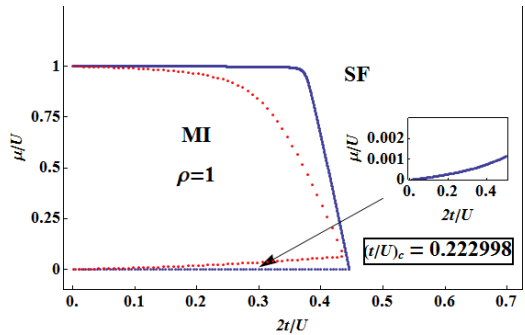}
\caption{(Color online) Phase diagram of the 1D extended BH model
with $\epsilon_{j}=0~\forall~j$ for $\rho=1$ and $M=10$ (blue solid line) and $M=6$ (red dotted line),
where the area surrounded by the phase lines denoted with ~`MI' stands for Mott-insulator
with density one, while the area outside the MI denoted  with `SF' is superfluid
phase, the inset shows the magnified lower phase boundary for $M=10$ case as a function of
$2t/U$, and the critical point value $(t/U)_{\rm c}$ shown is for the $M=10$ case.
The parameter $t_{0}=0.02597t$ when $M=6$ and $t_{0}=0.0002165t$ when $M=10$.}
\label{F1}
 \end{center}
 \end{figure}
\vskip -.5cm

{\bf Some ground state properties.}
At integer filling with $\rho=N/M$ being an integer,
the phase boundary in the $\mu$-$U$ ground-state
phase diagram of the original BH model may be determined by various
methods, e. g., those shown in [5-17].
The phase boundary may be determined
by the chemical potentials~\cite{[22]} defined by
$\mu^{+}(N,M,t_{0}/U)=E_{\rm g}(N+1,M,t_{0}/U)-E_{\rm g}(N,M,t_{0}/U)$
and
$\mu^{-}(N,M,t_{0}/U)=E_{\rm g}(N,M,t_{0}/U)-E_{\rm g}(N-1,M,t_{0}/U)$,
where $E_{\rm g}(N,M,t_{0}/U)$ is the ground state energy
of the model with $M$ sites and $N$ particles.
As an example of the application,
we show the ground state
phase diagram for the first Mott lobe ($\rho = 1$) obtained based on
the exact solutions (\ref{BA}), (\ref{BAC}), and (\ref{BAE}) for $U\neq 0$ and $\epsilon_{j}=0~\forall~j$
with $M=10$ sites in Fig. 2.
The critical point $(t/U)_{c}$ is determined by the condition that
$\delta(N,M,(t_{0}/U)_{c})=\mu^{+}(N,M,(t_{0}/U)_{c})-\mu^{-}(N,M,(t_{0}/U)_{c})=0$.
Since $t_{0}=2tN/D(N,M)$
is a very small quantity, the parameter $t$ is adjusted
to the sixth decimal place in order to get $\delta(N,M, (t_{0}/U)_{c})=0$ with error less than $10^{-7}$,
from which we get $(t/U)_{c}=0.222998$.
This critical value is smaller than that in the original 1D BH model,
in which  $(t/U)_{c}\sim 0.3$ as reported in \cite{[6],[10],[11],[13],[22],[23]}.
The overall shape of the Mott lobe of $\rho=1$ is also different
from that in the original 1D HB model.
The upper (lower)
phase line in the original BH model is  lower (upper) convex curve, while
the upper (lower) phase line in the extended BH model is upper (lower) convex curve.
Moreover, the upper phase boundary gradually lowers down with the increasing
of $t/U$ when $t/U\leq0.186$, while it lowers down drastically
with the increasing of $t/U$ when $t/U\geq0.186$.
The lower phase boundary gradually moves up with the increasing
of $t/U$ as shown in the inset of Fig. 2.
Since the results are obtained for finite number of sites,
as a comparison, the phase diagram of the model under the same
condition for $M=6$ sites is also shown in Fig. 2, which
indicates that the transitional behavior is enhanced
with the increasing of the number of sites.
Moreover, the critical point $(t/U)_{\rm c}$ is size-dependent.
Our calculation show that $(t/U)_{\rm c}\sim 0.218$ when $M=6$,
and $(t/U)_{\rm c}\sim 0.2243$ when $M=12$, namely
the larger the number of sites,  the slightly greater the
$(t/U)_{\rm c}$ value, though we can not figure out
the exact $(t/U)_{\rm c}$ value in the large-$M$ limit.

The probabilities to detect $n$ particles on a single site of the model with $M=6$ and
 $M=10$ sites and $\epsilon_{j}=0~\forall~j$  as functions
of $U/(2t)$ at zero temperature for $\rho=1$  are also calculated, of which the results are shown in the left panels of Fig. 3.
Contrary to the original 1D BH model~\cite{[24]}, where there is no
critical behavior in $P_n$, there is drastic change in $P_{n}$ for $n=0,~1,~2$
in the extended BH model, especially when $M$ is getting larger,
which occurs coincidentally at $U/t\sim 3.125<(U/t)_{c}=4.48435$ in these three $P_{n}$ curves
when $M=10$.
Moreover, these $P_{n}$ curves for the $M=10$ case already plateau in the SF regime with $U/t<(U/t)_{c}$
in contrast to the original BH model~\cite{[24]}, where these curves only plateau in the Mott
regime with $U/t\gtrsim (U/t)_{c}$.
It is will known that the critical behavior in $P_{n}$ is driven by
the quantum phase transition (QPT), while the critical point in
the $\mu$-$U$ phase diagram is determined by the condition
that $\delta(N,M,(t_{0}/U)_{c})=0$.
However, it is often expected that the position of the critical behavior in $P_{n}$
should be near or at $(U/t)_{c}$.
On the contrary, it seems that the
position of the critical
behavior in $P_{n}$ in the large-$M$ case always deviates away from
the critical point $(U/t)_{c}$ of the SF-MI transition in the extended model,
though it can not be verified in the large-$M$ limit at present.

To show the QPT in the ground state,
we calculate the simple mode entanglement measure proposed in \cite{[25]} with

\begin{equation}\label{ent}\eta=-{1\over{M}}\sum_{i=1}^{M}
{\rm Tr}\left\{(\phi)_{i}{\rm Log}_{N+1}(\phi)_{i}\right\},
\end{equation}
where $(\phi)_{i}$ ($i=1,\cdots,M$) is  {the} reduced density matrix of the ground state
with bosons on the $i$-th site only, which is the model-independent.
The measure (\ref{ent}) in the present case can further be simplified  as
\begin{equation}\label{ent2}\eta=-{\rm Tr}\left\{\phi{\rm Log}_{N+1}\phi\right\}
=-\sum_{n=0}^{N}P_{n}{\rm Log}_{N+1}P_{n},
\end{equation}
where $\phi=(\phi)_{i}$ for any $i$ because the ground state
is symmetric with respect to the site-permutations. (\ref{ent2}) indicates
that the probability to detect $n$ particles on a single site
$P_{n}$ is just the $n$-th diagonal matrix element of the reduced density matrix
$\phi=(\phi)_{i}$ for any $i$ in this case. Hence, the entanglement measure
(\ref{ent2}) may be practically useful since  $P_{n}$
are measurable experimentally~\cite{[24],[26],[27],[28]}.
It is shown in the right panels of Fig. 3 that the position of the drastic change in $\eta$
coincides with that of $P_{n}$ shown in the left panel for both the $M=6$ and the $M=10$ case,
which deviates away from $(U/t)_{c}$ of the SF-MI transition in the $M=10$ case,
especially when $M$ is getting larger.

\begin{figure}[H]
\begin{center}
\centerline{\hbox{\epsfig{file=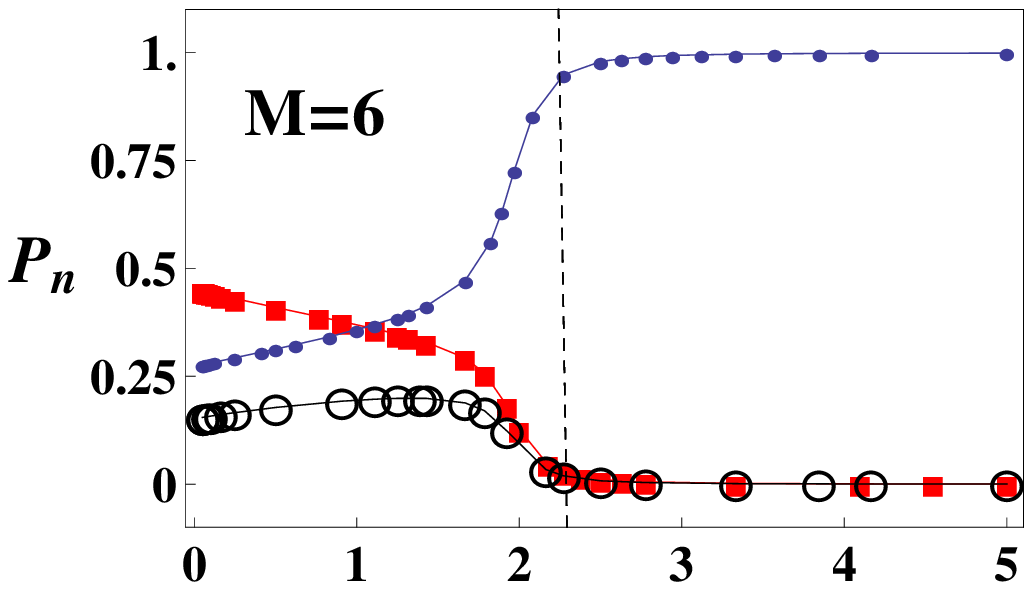, width=4.cm}~~~\epsfig{file=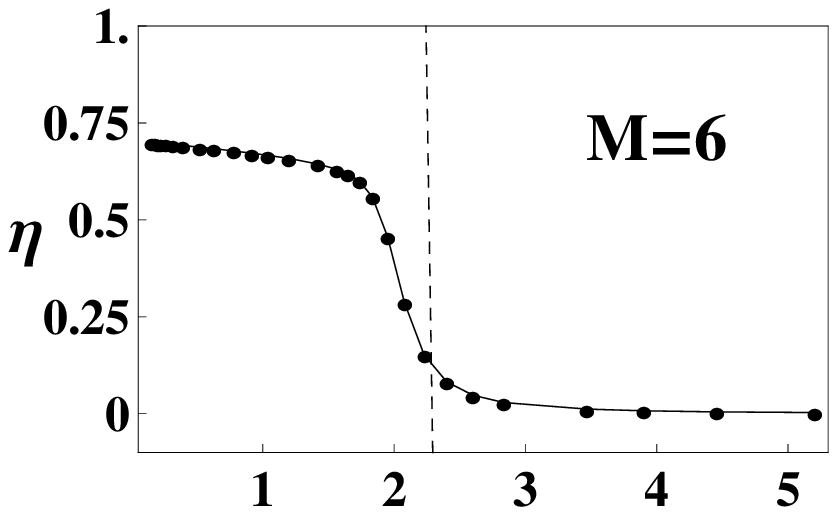, width=4cm}}}
\centerline{\hbox{~\epsfig{file=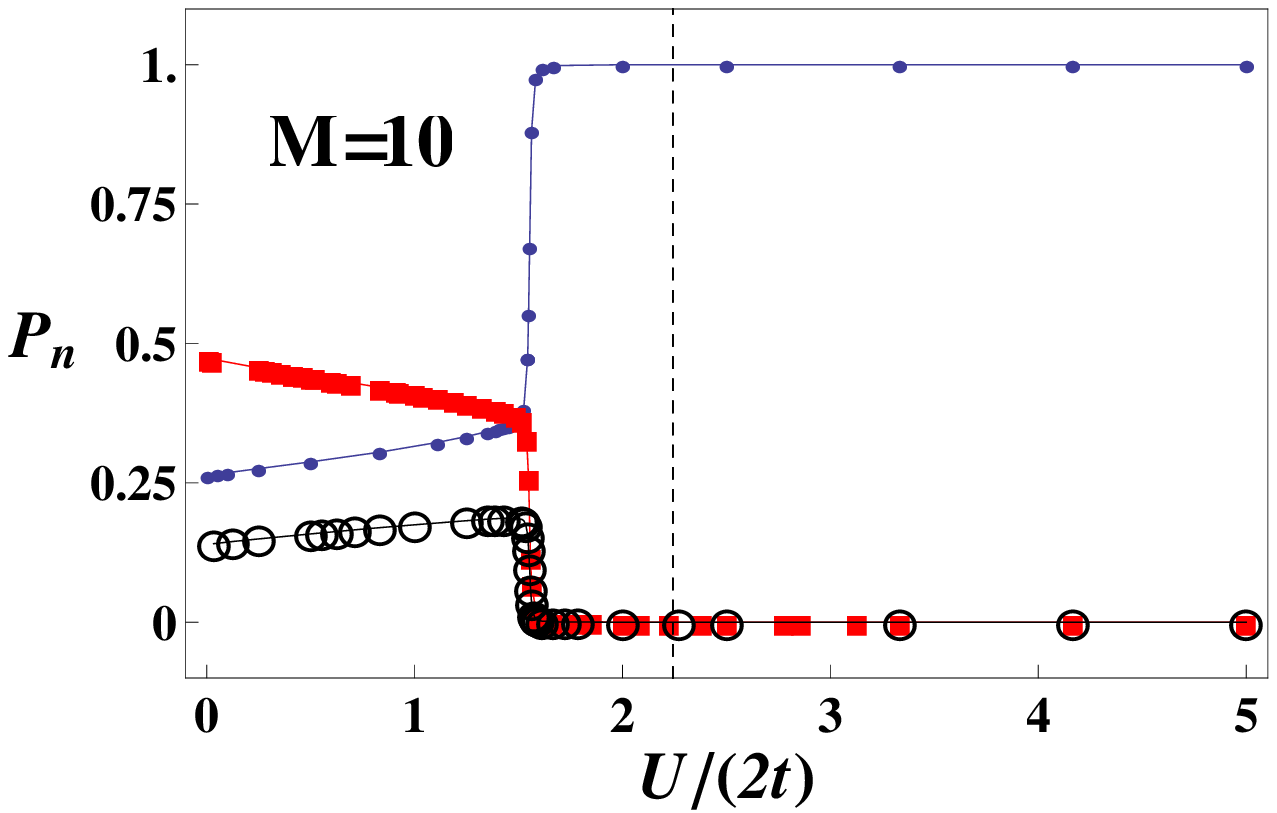, width=4.cm}~\epsfig{file=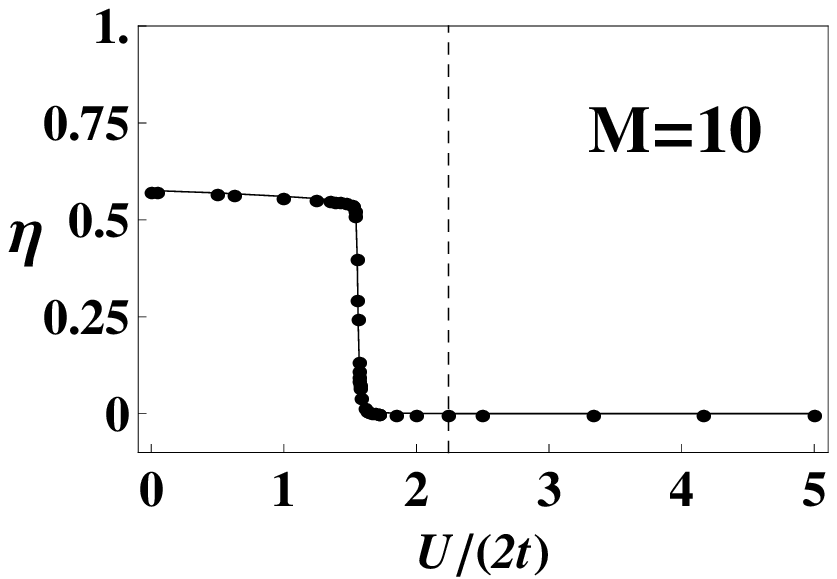, width=4cm}}}
\caption{(Color online)
The probabilities to detect $n$ particles on a single site (left panels) of the model
with $M=6$ and $M=10$ sites and $\epsilon_{j}=0~\forall~j$
as functions
of $U/(2t)$ at zero temperature for $\rho=1$,
where $P_{0}$, $P_{1}$,
and $P_{2}$ are shown by solid squares, solid dots, and open circles, respectively,
the vertical dashed line indicates the critical point position of
the SF-MI transition. Ground state entanglement measure $\eta$ (right panels) of the model as
a function of $U/(2t)$ under the same conditions as those for the left panels.
} \end{center}
 \end{figure}

{\bf Conclusions.}
In this work, an extended Bose-Hubbard (BH) model with number-dependent multi-site and  
infinite-range hopping is proposed, which, similar to the original BH model, 
describes a phase transition between the delocalized
superfluid (SF) phase and localized Mott insulator (MI) phase.
It is shown that the model with local Euclidean $E_{2}$ symmetry 
is exactly solvable when on-site local potential or disorder is included,
while the model without local potential or disorder is quasi-exactly solvable, 
which means only a part of the excited states including the ground state being exactly solvable.
As applications of the exact solution for the ground state,
phase diagram of the model in 1D without local potential and on-site disorder for filling factor $\rho=1$
with  $M=6$ sites and that with $M=10$  are obtained.
The probabilities to detect $n$ particles on a single site, $P_{n}$, for $n=0,~1,~2$  as functions
of the control parameter $U/t$  in these two cases are also calculated. It is shown that
the critical point in $P_{n}$ and in the entanglement measure is away from that of 
the SF-MI transition determined in the phase analysis in the finite-$M$ cases exemplified.
In order to verify whether the critical point in $P_{n}$ is also away
from that of the SF-MI transition determined in the phase analysis in the large-$M$
limit, large scale calculations should be carried out, which will be a part of our
future work. It is also shown that the model-independent entanglement measure
is related with $P_{n}$, which, therefore, may be practically useful
because $P_{n}$ is measurable experimentally. 
\\

{\bf Acknowledgement.} Support from the U.S. National Science Foundation
(OCI-0904874), the Southeastern Universities Research Association, the
Natural Science Foundation of China (11175078; 11375080),  and the LSU--LNNU joint
research program (9961) is acknowledged.

\end{document}